\documentclass[prd, aps, nobibnotes, nofootinbib,
eqsecnum,
preprint,
showpacs,
preprintnumbers,
amsmath,
amssymb]{revtex4-1}

\usepackage{graphicx}
\usepackage{bm}
\usepackage{hyperref}
\usepackage{mathrsfs}
\usepackage{xcolor}
\usepackage{subcaption}
\usepackage{multirow}
\usepackage{hhline}
\usepackage{comment}

\begin{document}

\title{{\bf{\Large Newtonian cosmology from quantum corrected Newtonian potential }}}

\author{ Rituparna Mandal}
\email{drimit.ritu@gmail.com, rituparna1992@bose.res.in}\,
\affiliation{Department of Astrophysics and High energy physics, S.N. Bose National Centre for Basic Sciences, Block JD, Sector III, Salt Lake, Kolkata 700106, India}
\author{ Sunandan Gangopadhyay}
\email{sunandan.gangopadhyay@gmail.com, sunandan.gangopadhyay@bose.res.in}\,
\affiliation{Department of Astrophysics and High energy physics, S.N. Bose National Centre for Basic Sciences, Block JD, Sector III, Salt Lake, Kolkata 700106, India}
\author{ Amitabha Lahiri}
\email{amitabha@boson.bose.res.in }
\affiliation{Department of Astrophysics and High energy physics, S.N. Bose National Centre for Basic Sciences, Block JD, Sector III, Salt Lake, Kolkata 700106, India}

\begin{abstract}
We study the Newtonian cosmology taking into account the leading classical and quantum corrections of order $\mathcal{O}(G^{2})$ in the Newtonian potential. We first derive the modified Friedmann equations starting from the non-relativistic conservation of kinetic energy and potential energy for an infinitesimal mass. We then consider the leading classical correction term and the quantum correction term in the Newtonian potential for deriving the Friedmann equation, however, the quantum correction term is too small and hence does not contribute in the physical results. We then investigate the difference in scale factor with the usual scale factor for various matter like radiation, dust and cosmological constant by considering the corrections in the Newtonian potential. We observe that the evolution of the universe is similar for radiation and dust cases at late times. The cosmological constant case shows a steep increase in the scale factor compared to the other cases. We also note that the universe may have a bounce in the case of radiation depending on the sign of the coefficient of the leading classical correction.
\end{abstract}
\maketitle

\newpage
\section{Introduction}
The standard model of cosmology is based upon the Friedmann equation, which essentially describes a relativistic, homogeneous and isotropic universe. It is derived from the general theory of relativity and its solution provides the scale factor, the rate of expansion, and various derived quantities. Somewhat surprisingly, it is also possible to derive the Friedmann equation using only Newtonian mechanics and Newtonian gravitation, and it turns out that locally observable predictions are identical in the two cases~\cite{Milne 1934, McCrea 1934, W McCrea 1955, McCrea 1955, Callan_2005, Jordan 2005}. This approach starts from the conservation of  kinetic energy and gravitational potential energy for the motion of a galaxy at the surface of a very large sphere. The important ingredient in this calculation is the static Newtonian potential. Since that is only useful when pressure vanishes, the original derivations were only for a Friedmann universe filled with pressureless dust. However it can be shown, either by reducing from a general relativistic treatment or by intuitive (but general relativistic) arguments, that a similar treatment can hold for other types of matter (with $p\neq0$) including radiation~\cite{kour}. Then the Newtonian approach becomes especially valuable, as it shows the simple nature of the cosmological equations and brings out the essential characteristics of relativistic cosmology bypassing the complexity of the underlying mathematics.


Recently, modified Friedmann equations were obtained in~\cite{Bargueno 2016} by considering  quantum corrected Newtonian potential of two non-relativistic masses. Such quantum corrections have been there in the literature and arise from quantum loop calculations~\cite{Donoghue 1994, Hamber 1995, Muzinich 1995, Akhundov 1998, Khriplovich 2002, Bjerrum 2003, Kirilin 2007, Donoghue 2012, Bjerrum 2015, Wang 2015, Burgess 2004, Duff 1974, Radkowski 1970}. The interesting feature of such results is that the Newtonian potential gets a quantum correction along with an $\mathcal{O}(\hbar^{0})$ classical term  which surprisingly comes out from a loop calculation. These quantum corrections appearing in the Newtonian potential originate due to the non-analytic pieces of the one loop amplitude considering the lowest order Einstein action. There has been some disagreement in the literature about the coefficient of the leading order term due to the existence of an unavoidable ambiguity in defining the potential~\cite{Bjerrum 2003}. Effects of the corrections to the Newtonian potential have also been investigated in the three-body problem of classical celestial mechanics \cite{battista1}-\cite{battista4}. It is thus natural to consider the Newtonian approach to Friedmann cosmology if we wish to find the effect of the corrections arising from loop calculations. 

Our first aim is to derive the corrected Friedmann equation including both the leading classical and quantum corrections to the Newtonian potential. To study the cosmology further, we need another equation which accounts for the work done by the pressure as the universe expands. This is called the continuity equation. With the help of these two equations, we investigate the solutions of the scale factors taking into account the corrections terms through Friedmann equation for different kinds of matter like radiation, dust and cosmological constant. 
We note here that the quantum correction term in the Friedmann equation is too small to make any difference in the observed result, so we mainly concentrate on the contribution of the classical correction term to the Friedmann equation. We also observe that the sign of the coefficient of the classical term is crucial, the negative sign of the coefficient can give rise to a bounce in case of radiation no matter what the signature of the coefficient of the quantum correction term may be. We have considered a positive sign of the coefficient in the classical correction \cite{Donoghue 1994}. In this case the bounce does not appear.

The paper is organized as follows. In section 2, we describe the set up for the Friedmann equation from Newtonian dynamics. In section 3, we derive the modified Friedmann equation taking into account the classical and quantum corrections in the Newtonian potential. In section 4, we study the cosmology and calculate the corrected scale factor for different matter like radiation, dust and cosmological constant. We conclude in section 5.
   
\section{Friedmann equation from Newtonian dynamics}
In this section, we first briefly review the derivation of the Friedmann equations from Newtonian gravity \cite{Wienberg, Jordan 2005, Bargueno 2016}. To proceed, one needs to consider the expansion of the universe. This is done by writing 
\begin{align}
\frac{\mathrm{d}R(t)}{\mathrm{d}t}=HR(t)
\label{hubble}
\end{align}
where $H$ is the Hubble parameter and $R(t)$ is the radius of the universe. Now for an object, for example a galaxy of mass $m$ at the surface of the sphere of radius $R$ with mass $M$ which is uniformly distributed, the total energy is given by
 \begin{align}
E=\frac{1}{2}m\left(\frac{\mathrm{d}R}{\mathrm{d}t} \right)^{2}-\frac{GMm}{R}
 \label{energy}
 \end{align}
where the first term on right hand side represents the kinetic energy and the second term represents the potential energy of the mass $m$ at the surface of the sphere of radius $R$ which has a mass $M$ of the universe. Simplifying the above equation, we obtain 
\begin{align}
\frac{2E}{mR^{2}}=\frac{1}{R^{2}}\left(\frac{\mathrm{d}R }{\mathrm{d} t}\right)^{2}-\frac{2GM}{R^{3}}~.
\label{e1}
\end{align}
The mass of the universe which is a sphere of radius $R$ is given by
\begin{align}
M=\frac{4}{3}\pi R^{3}\rho
\label{mass}
\end{align}
where $\rho$ is the energy density of the universe. 

\noindent Multiplying eq.(\ref{e1}) by $R^2$ leads to a form which expresses the law of conservation of energy and follows by integrating the Newtonian equation of motion of an unit mass of cosmic fluid situated at a distance $R(t)$ from a point where the entire mass $M$ of the universe is concentrated. The equation is fundamental because general relativity shows that it is much more general than the Newtonian equation of motion that holds only with the assumption of the pressure of the cosmic fluid being zero \cite{kour}.
 
\noindent Substituting eq.\eqref{mass} in eq.\eqref{e1} and defining $H=\frac{1}{R}\left(\frac{\mathrm{d}R }{\mathrm{d} t}\right)$, we get 
\begin{align}
\frac{2E}{mR^{2}}=H^{2}-\frac{8}{3}\pi G \rho~.
\label{e2}
\end{align}
Since $E$ and $m$ are constants, we can define a new constant $K=- \frac{2E}{m}$ which gives
\begin{align}
H^{2}= \frac{8}{3}\pi G \rho -\frac{K}{R^{2}}~.
\label{friedmann}
\end{align}
This equation is the first Friedmann equation which has now been derived using Newtonian gravity \cite{Wienberg}.\\
Now we need to get another equation to solve $R$ and $\rho$ as a function of time. For this, we write the first law of thermodynamics
\begin{align}
\mathrm{d}Q=\mathrm{d}U+p\mathrm{d}V
\label{first}
\end{align}
where $\mathrm{d}Q$ is the heat flow into or out of the volume, $\mathrm{d}U$ is the change in internal energy, $p$ is the pressure and $\mathrm{d}V$ is the change in the volume. Taking $\mathrm{d}Q=0$ for homogeneous and isotropic universe, the first law of thermodynamics for the expanding universe reduces to
\begin{align}
\mathrm{d}\left(\rho \frac{4}{3}\pi R^{3}\right)+p\mathrm{d}\left(\frac{4}{3}\pi R^{3}\right)=0~.
\label{thermo}
\end{align}    
Here the proper volume of the sphere is $ V=\frac{4}{3}\pi R^{3}$ and the internal energy of the sphere takes the form as $U=\rho \frac{4}{3}\pi R^{3}$.  Now simplifying the above equation, we get the wellknown conservation law
\begin{align}
 R\frac{\mathrm{d} \rho}{\mathrm{d} t}+3\left(p+\rho \right)\frac{\mathrm{d}R }{\mathrm{d} t}=0~.
\label{cons}
\end{align}
The first Friedmann equation, multiplied by $R^{2}$, takes the form
\begin{align}
\left(\frac{\mathrm{d}R}{\mathrm{d}t}\right)^{2}= \frac{8}{3}\pi G \rho R^{2} -K~.
\label{modified}
\end{align}  
Now taking the derivative of the above equation with respect to $t$ and replacing $R \frac{\mathrm{d}\rho}{\mathrm{d}t}$ from the energy conservation (eq.\eqref{cons}), one arrives at the second Friedmann equation
\begin{align}
\frac{\mathrm{d}^{2}R }{\mathrm{d} t^{2}}=-\frac{4 \pi G}{3}\left (3 p+\rho \right )R~. 
\label{friedmann2}
\end{align}


\section{Quantum corrected Friedmann equations}
\noindent In this section, we investigate the quantum corrected Friedmann equations taking into account the leading quantum correction of the Newtonian potential. These corrections have been obtained by treating gravity as an effective field theory \cite{Donoghue 1994, Khriplovich 2002, Bjerrum 2003, Kirilin 2007, Donoghue 2012, Burgess 2004}. In~\cite{Donoghue 1994, Bjerrum 2003}, it was shown that the Fourier transform of the one-loop scattering amplitude of two masses gives a quantum correction to the Newtonian potential. 
The corrected Newtonian potential reads \cite{Bjerrum 2003}
\begin{eqnarray}
V(r)=-\frac{GMm}{r} \left[1+\lambda\frac{G(M+m)}{rc^{2}}-\zeta\frac{ G \hbar}{c^3 r^2} +...\right]
\label{1}
\end{eqnarray}
where $\lambda$ and $\zeta$ are parameters which take the value $\lambda=3$ and $\zeta=-\frac{41\pi}{10}$ obtained using the full scattering amplitude as the definition of the nonrelativistic potential \cite{Bjerrum 2003}. 
The $\mathcal{O}(\frac{G^{2}}{r^2})$ and $\mathcal{O}(\frac{l_{Pl}^{2}}{r^3})$ terms have come from loop corrections to the Newtonian potential considering gravity as an effective theory \cite{Donoghue 2012}. It is surprising to note that $\mathcal{O}(\frac{G^{2}}{r^2})$ is a classical term coming from loop correction. This was pointed out in \cite{Iwasaki}. 

Now we will proceed to derive the quantum corrected Friedmann equations. Once again starting with the energy conservation and taking the correction terms in the Newtonian potential, we have
\begin{align}
E=\frac{1}{2}m\left(\frac{\mathrm{d}R }{\mathrm{d} t}\right)^{2}-\frac{GMm}{R}-\lambda\frac{G^{2}Mm(M+m)}{R^{2}c^{2}}+\zeta \frac{GMm}{R}\frac{ l_{Pl}^{2}}{R^{2}}~.
\label{4}
\end{align}
Rewriting the above equation, we obtain
\begin{align}
\frac{2E}{mR^{2}}=\frac{1}{R^{2}}\left(\frac{\mathrm{d}R }{\mathrm{d} t}\right)^{2}-\frac{2GM}{R^{3}}-2\lambda\frac{G^{2}M^{2}}{R^{4}c^{2}}\left(1+\frac{m}{M}\right)+\zeta \frac{2GM}{R^{3}}\frac{ l_{Pl}^{2}}{R^{2}}~.
\label{5}
\end{align}
Neglecting the term of $\mathcal{O}\left(\frac{m}{M} \right)$, the above equation simplifies to  
\begin{align}
\frac{2E}{mR^{2}}=\frac{1}{R^{2}}\left(\frac{\mathrm{d}R }{\mathrm{d} t}\right)^{2}-\frac{2GM}{R^{3}}-2\lambda\frac{G^{2}M^{2}}{R^{4}c^{2}}+\frac{2GM}{R^{3}}\frac{\zeta l_{Pl}^{2}}{R^{2}}~.
\label{6}
\end{align}
Substituting the mass $M=\frac{4}{3}\pi R^3 \rho$, and defining $H=\frac{1}{R}\left(\frac{\mathrm{d}R }{\mathrm{d} t}\right)$, the above equation reads
\begin{align}
\frac{2E}{mR^{2}}=H^{2}-\frac{8}{3}\pi G \rho-\frac{32 \lambda}{9c^2}\pi^{2}R^{2} G^{2} \rho^{2}+ \frac{8}{3}\pi G \rho\zeta \left(\frac{l_{Pl}}{R}\right)^{2} ~. 
\label{7}
\end{align}
Defining $K=-\frac{2E}{m}$, we find the following equation 
\begin{align}
H^{2}= \frac{8}{3}\pi G \rho -\frac{K}{R^{2}}+\frac{32 \lambda}{9c^2}\pi^{2}R^{2} G^{2} \rho^{2}- \frac{8}{3}\pi G \rho\zeta \left(\frac{l_{Pl}}{R}\right)^{2} ~. 
\label{8}
\end{align}
This is the first Friedmann equation taking into account the corrections terms in the Newtonian potential. Now using the energy conservation eq.\eqref{cons} and the quantum corrected first Friedmann equation, we get the quantum corrected second Friedmann equation 
 \begin{align}
 \frac{\mathrm{d}^{2}R }{\mathrm{d} t^{2}}=-\frac{4 \pi G}{3}\left ( 3p+\rho \right )R-\frac{32 \lambda}{9c^2}\pi^{2}R^{3} G^{2} \rho\left ( 3p+\rho \right )+4\pi G\zeta l_{Pl}^{2}\frac{\left ( p+\rho \right )}{R}~.
 \label{qfri}
 \end{align}

\section{Quantum corrected Newtonian Universe}
To proceed further, we can see that we have two independent equations out of the three equations, namely, the first  Friedmann equation, second Friedmann or acceleration equation and the continuity equation.
Thus we have a system of  two independent equations with three unknowns $R(t)$, $\rho$ and $p$. Hence to solve this system of equations, we need an equation of state
\begin{align}
p=\left(\gamma-1\right)\rho~,~~\gamma=\text{constant}~.
\label{9}
\end{align}
Now we can solve eq.(s)(\ref{8}, \ref{cons}, \ref{9}) to obtain $\rho(t)$, $p(t)$ and $R(t)$ for all times. In reality, our universe can contain different components with different equations of state during the evolution of the universe. 
Fortunately, the energy density and pressure for the different components of the universe are additive. The continuity equation separately holds for each component as there is no interaction between them. In this work, we will study the cosmology for a single component. Solving the continuity eq.\eqref{cons}, we obtain $\rho$ in terms of $R$ 
 \begin{align}
\rho=\rho_{0}\left(\frac{R_{0}}{R}\right)^{3\gamma}
\label{10}
\end{align}
where $\rho_{0} \equiv \rho(R_{0})$ is the energy density at the present time, with $R_{0}$ being the present radius of the universe.\\
Inserting this into the quantum corrected first Friedmann equation (eq.\eqref{8}) with $K=0$, we get
\begin{align}
\frac{1}{R^{2}} \left(\frac{\mathrm{d} R}{\mathrm{d} t}\right)^{2}=\frac{8}{3}\pi G \rho_{0}\left(\frac{R_{0}}{R}\right)^{3\gamma}+\frac{32 \lambda}{9c^2}\pi^{2}R^{2} G^{2} \rho_{0}^{2}\left(\frac{R_{0}}{R}\right)^{6\gamma}- \frac{8}{3}\pi G \rho_{0}\zeta\left(\frac{R_{0}}{R}\right)^{3\gamma}  \left(\frac{l_{Pl}}{R}\right)^{2}~.
\label{11}
\end{align}
Defining $\phi=\frac{4}{3 c^2}\pi G \rho_{0} R_{0}^{2}$, we can write the above equation in a simplified form as
\begin{align}
\left(\frac{\mathrm{d} R}{\mathrm{d} t}\right)^{2}=2c^{2}\phi\left(\frac{R_{0}}{R}\right)^{3\gamma-2}+2c^{2}\lambda\phi^{2}\left(\frac{R_{0}}{R}\right)^{6\gamma-4}-2\zeta c^{2} \phi  \left(\frac{R_{0}}{R}\right)^{3\gamma}\left(\frac{l_{Pl}}{R_{0}}\right)^{2}~.
\end{align}
Simplifying the above equation, we finally reach
\begin{align}
\left(\frac{\mathrm{d} R}{\mathrm{d} t}\right)^{2}=2c^{2}\phi\left(\frac{R_{0}}{R}\right)^{3\gamma-2}\left[1+\lambda \phi \left(\frac{R_{0}}{R}\right)^{3\gamma-2}- \zeta \left(\frac{l_{Pl}}{R_{0}}\right)^{2} \left(\frac{R_{0}}{R}\right)^{2}\right] ~.
\label{11}
\end{align}
We will now proceed to solve the above equation for various matter like  radiation $(\gamma=\frac{4}{3})$, dust $(\gamma=1)$ and cosmological constant $(\gamma=0)$. 
\subsection{Radiation $(\gamma=\frac{4}{3})$} 
\noindent We will now study spatially flat universe containing only radiation.
Setting $\gamma=\frac{4}{3}$ in eq.\eqref{11}, we obtain the following differential equation
\begin{align}
\left(\frac{\mathrm{d} R}{\mathrm{d} t}\right)^{2}&=2c^{2} \phi \left(\frac{R_{0}}{R}\right)^{2}
 \left[1+\left(\lambda \phi - \zeta \left(\frac{l_{Pl}}{R_{0}}\right)^{2} \right)\frac{R_{0}^{2}}{R^{2}}\right] \nonumber  \\ & \equiv a_{1}\left(\frac{R_{0}}{R}\right)^{2}+a_{2}\left(\frac{R_{0}}{R}\right)^{4}~.
\label{12}
\end{align}
Here we have defined $a_{1}=2c^{2} \phi$ and $a_{2}=2c^{2} \phi \left(\lambda \phi - \zeta \left(\frac{l_{Pl}}{R_{0}}\right)^{2} \right)$ for convenience. \\
Using eq.\eqref{10} with $\gamma=\frac{4}{3}$, we can recast the first Friedmann equation for the radiation dominated, spatially flat universe in terms of density $\rho$ as
\begin{align}
H^{2}=\frac{8}{3}\pi G \rho \left[1+\left(\frac{\rho}{\rho_{0}}\right)^{1/2}\left(\lambda \phi-\frac{\zeta l_{Pl}^{2}}{R_{0}^{2}}\right)\right]~.
\label{12a}
\end{align}
It can be seen easily that the sign of the second term is fixed by $\lambda \phi$ which is the dominant term.
Hence, the Hubble parameter can vanish for a particular density depending on the sign of $\lambda$. For $\lambda <0$, $H$ can vanish. However, for $\lambda >0$ the Hubble parameter can never vanish for any value of $\rho$ since $\frac{\zeta l_{Pl}^{2}}{R_{0}^{2}}$ is negligibly small compared to $\lambda \phi$.       

Solving for the radius of the universe $R(t)$ from eq.\eqref{12}, we get 
%
%
%
\begin{align}
\sqrt{a_{1}}\frac{t}{R_{0}}=\frac{R}{2R_{0}}\left(\frac{R^{2}}{R_{0}^{2}}+ \frac{a_{2}}{a_{1}} \right)^{\frac{1}{2}}-\frac{a_{2}}{2a_{1}} \ln\left[\sqrt{\frac{a_{1}}{a_{2}}}\left(\frac{R}{R_{0}}+\left(\frac{R^{2}}{R_{0}^{2}}+ \frac{a_{2}}{a_{1}} \right)^{\frac{1}{2}}\right)\right]+B
\label{ncos.15}
\end{align}
where $B$ is an integration constant. Setting the initial condition $R(t_{0})=R_{0}$  at present time $t_{0}$, gives 
\begin{align}
B=\sqrt{a_{1}}\frac{t_{0}}{R_{0}}-\frac{1}{2}\sqrt{1+ \frac{a_{2}}{a_{1}}}+\frac{a_{2}}{2a_{1}} \ln\left[\sqrt{\frac{a_{1}}{a_{2}}}\left(1+\sqrt{1+ \frac{a_{2}}{a_{1}}}\right)\right]~.
\label{conB}
\end{align}
The solution in terms of the scale factor $a(t)=\frac{R(t)}{R_{0}}$ therefore takes the form
\begin{align}
\sqrt{a_{1}}\frac{\left(t-t_{0}\right)}{R_{0}}=\frac{1}{2}\left[a\sqrt{a^{2} +\frac{a_{2}}{a_{1}}}-\sqrt{1+ \frac{a_{2}}{a_{1}}}\right]-\frac{a_{2}}{2a_{1}} \ln\left[\frac{a+\sqrt{a^{2}+ \frac{a_{2}}{a_{1}}}}{1+\sqrt{1+ \frac{a_{2}}{a_{1}}}}\right]~.
\label{ncos.16}
\end{align}
Setting $\lambda=0$ and $\zeta=0$, the classical solution for the radiation dominated universe is recovered
\begin{align}
a=\sqrt{2\tau+1}
\label{classrad}
\end{align}
where $\tau=\frac{\sqrt{a_{1}}}{R_{0}}(t-t_{0})$. The classical solution is identical to the one obtained from general relativity.

\begin{figure}[htbp]
\includegraphics[scale=.6]{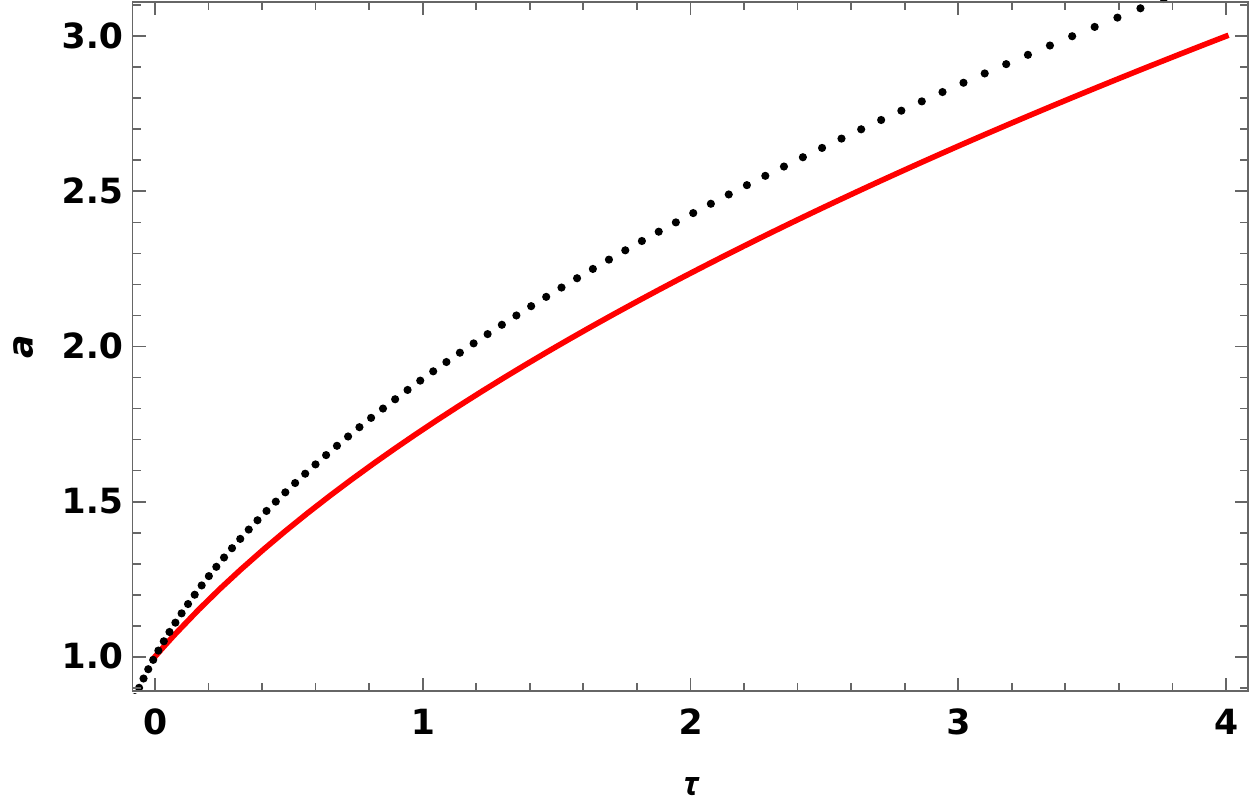}
\caption{Scale factor versus $\tau=\frac{\sqrt{a_{1}}}{R_{0}}(t-t_{0})$ for a radiation dominated universe. Solid red curve: classical scale factor. Dashed black curve:  corrected scale factor. }
\label{fig1}
\end{figure}

In Fig.\eqref{fig1}, we have plotted the scale factor $a$ for the Newtonian universe for both the classical and the corrected cases, with the dimensionless quantity $\tau$. As we have taken radiation as a single component, we should consider the density parameter $\Omega_{r,0}=\frac{\rho_{0}}{\rho_{c,0}}\approx 1$. $\rho_{c,0}$ is the critical mass density at present time, which is taken to be $\rho_{c,0}=9.2 \times 10^{-27}$  $kg~m^{3}$. Taking the Hubble constant $H_{0}=70$ $km~s^{-1}~Mpc^{-1}$, the present comoving radius of the universe is given by $R_{0}=1.3 \times 10^{26}~m$. Using these values, the value of $\phi$ comes out to be $0.48$. The value of the parameter $\lambda$ is taken as $\lambda=3$ \cite{Bjerrum 2003}. Hence the constant $\frac{a_{2}}{a_{1}}$ is given by $\frac{a_{2}}{a_{1}}=\left(\lambda \phi-\frac{\zeta l_{Pl}^{2}}{R_{0}^{2}}\right)\approx 1.44$.
%
\subsection{Dust $(\gamma=1)$}
\noindent In this subsection, we consider the matter to be dust. Setting $\gamma=1$ in eq.\eqref{11}, we obtain
\begin{align}
\left(\frac{\mathrm{d} R}{\mathrm{d} t}\right)^{2}&=2c^{2}\phi\frac{R_{0}}{R}\left[1+\lambda \phi  \frac{R_{0}}{R}- \zeta \left(\frac{l_{Pl}}{R_{0}}\right)^{2} \left(\frac{R_{0}}{R}\right)^{2}\right] \nonumber \\ & \equiv b_{1}\frac{R_{0}}{R}+b_{2}\left(\frac{R_{0}}{R}\right)^{2}+b_{3}\left(\frac{R_{0}}{R}\right)^{3}~
\label{19}
\end{align}
where $b_{1}=2c^{2}\phi$, $b_{2}=2c^{2}\lambda \phi^{2}$ and $b_{3}=-2 \zeta c^{2}\phi \left(\frac{l_{Pl}}{R_{0}}\right)^{2}$. 
%
Integrating the above equation, we get
\begin{align}
\sqrt{b_{1} R_{0}}t=\frac{2R_{0}^{3/2}}{3}\left(\frac{R}{R_{0}}-2 \frac{b_{2}}{b_{1}}\right)\left( \frac{R}{R_{0}}+ \frac{b_{2}}{b_{1}}\right) ^{\frac{1}{2}}+\frac{b_{3}R_{0}^{3/2}}{b_{1}}\frac{1}{\left(\frac{R}{R_{0}}+ \frac{b_{2}}{b_{1}}\right)^{\frac{1}{2}}}+C
\label{24}
\end{align}
where $C$ is an integration constant. Setting the initial condition $R(t_0)=R_0$ as before, we get 
\begin{align}
C=\sqrt{b_{1} R_{0}} t_{0}-\frac{2R_{0}^{3/2}}{3}\left(1-2 \frac{b_{2}}{b_{1}}\right)\left( 1+ \frac{b_{2}}{b_{1}}\right)^{\frac{1}{2}}-\frac{b_{3}}{b_{1}}\frac{R_{0}^{3/2}}{\left(1+ \frac{b_{2}}{b_{1}}\right)^{\frac{1}{2}}}~.
\label{intc}
\end{align}
The solution in terms of the scale factor therefore reads
\begin{align}
\frac{\sqrt{b_{1}}}{R_{0}} (t-t_{0})&=\frac{2}{3}\left(a-2 \frac{b_{2}}{b_{1}} \right)\left(a+ \frac{b_{2}}{b_{1}}\right) ^{\frac{1}{2}}-\frac{2}{3}\left(1-2 \frac{b_{2}}{b_{1}}\right)\left( 1+ \frac{b_{2}}{b_{1}} \right)^{\frac{1}{2}}+\frac{b_{3}}{b_{1}} \nonumber \\& \times \frac{R_{0}^{3/2}}{\left(\frac{R}{R_{0}}+ \frac{b_{2}}{b_{1}}\right)^{\frac{1}{2}}} -\frac{b_{3}}{b_{1}}\frac{R_{0}^{3/2}}{\left(1+ \frac{b_{2}}{b_{1}}\right)^{\frac{1}{2}}}~~.
\label{ncos.25}
\end{align}  
The classical solution from eq.\eqref{ncos.25} can be recovered by setting $\lambda=0$ and $\zeta=0$ and thus $b_{2}=b_{3}=0$, and reads
\begin{align}
a=\left(\frac{3}{2}\tau+1\right)^{\frac{2}{3}}
\label{classdust}
\end{align}
where $\tau = \frac{\sqrt{b_{1}}}{R_{0}} (t-t_{0}) $. As in the radiation case, this is identical with the solution obtained using general relativity. 

\begin{figure}[htbp]
\includegraphics[scale=.6]{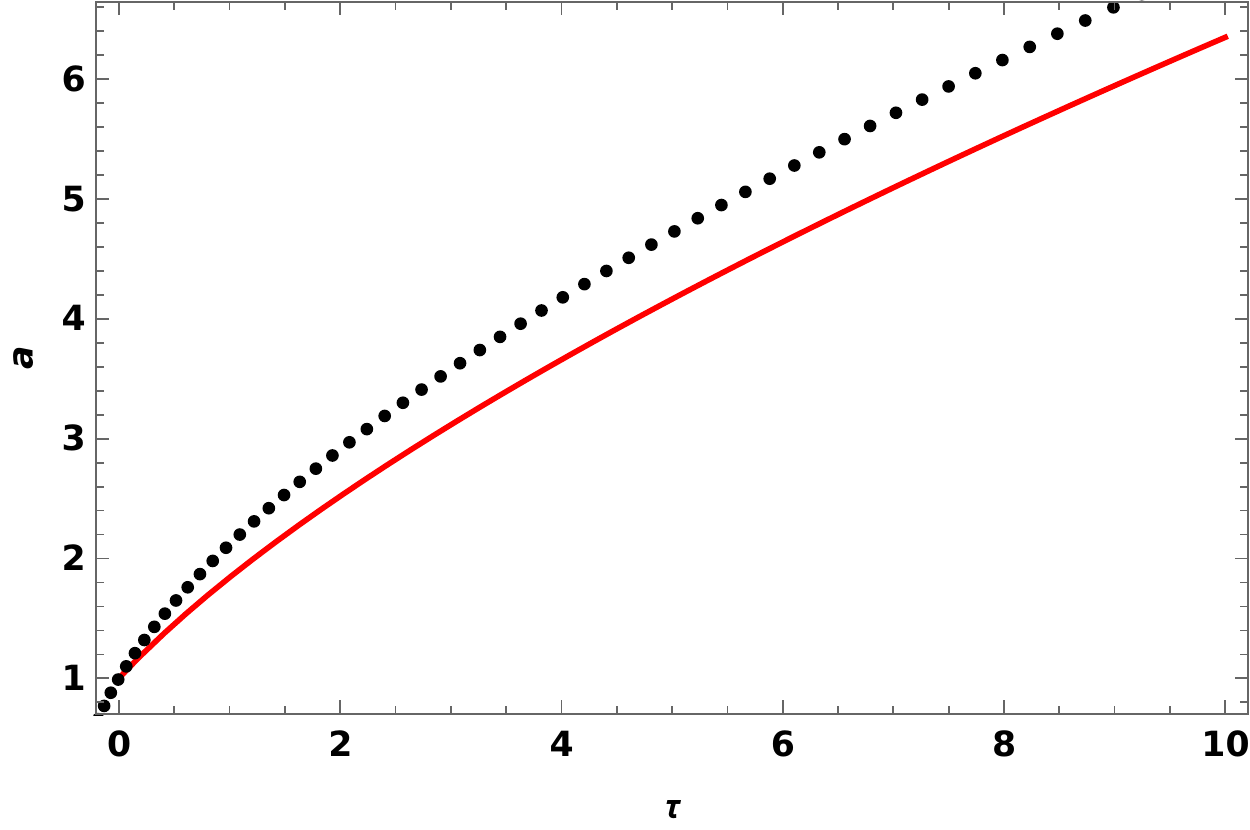}
\caption{Scale factor versus $\tau$ with dust. Solid red curve: classical scale factor. Dashed black curve: corrected scale factor. }
\label{fig2}
\end{figure}

In Fig.\eqref{fig2}, we have plotted the scale factor for spatially flat, dust dominated Newtonian universe for both the classical and the corrected cases. Here we have also used $\rho_{0}=\rho_{c,0}$ due to the fact that the density parameter for a universe with a single matter component 
is $\Omega_{m,0}=\frac{\rho_{0}}{\rho_{c,0}}=1$. So, the constant $\frac{b_{2}}{b_{1}}=\lambda \phi \approx 1.44$ as in the previous case. Another constant $\frac{b_3}{b_{1}}=-\zeta \left(\frac{l_{Pl}}{R_{0}}\right)^{2}$ is negligibly small and therefore does not contribute at the present time.
%
\subsection{Cosmological constant $(\gamma=0)$}
\begin{figure}[htbp]
\includegraphics[scale=.6]{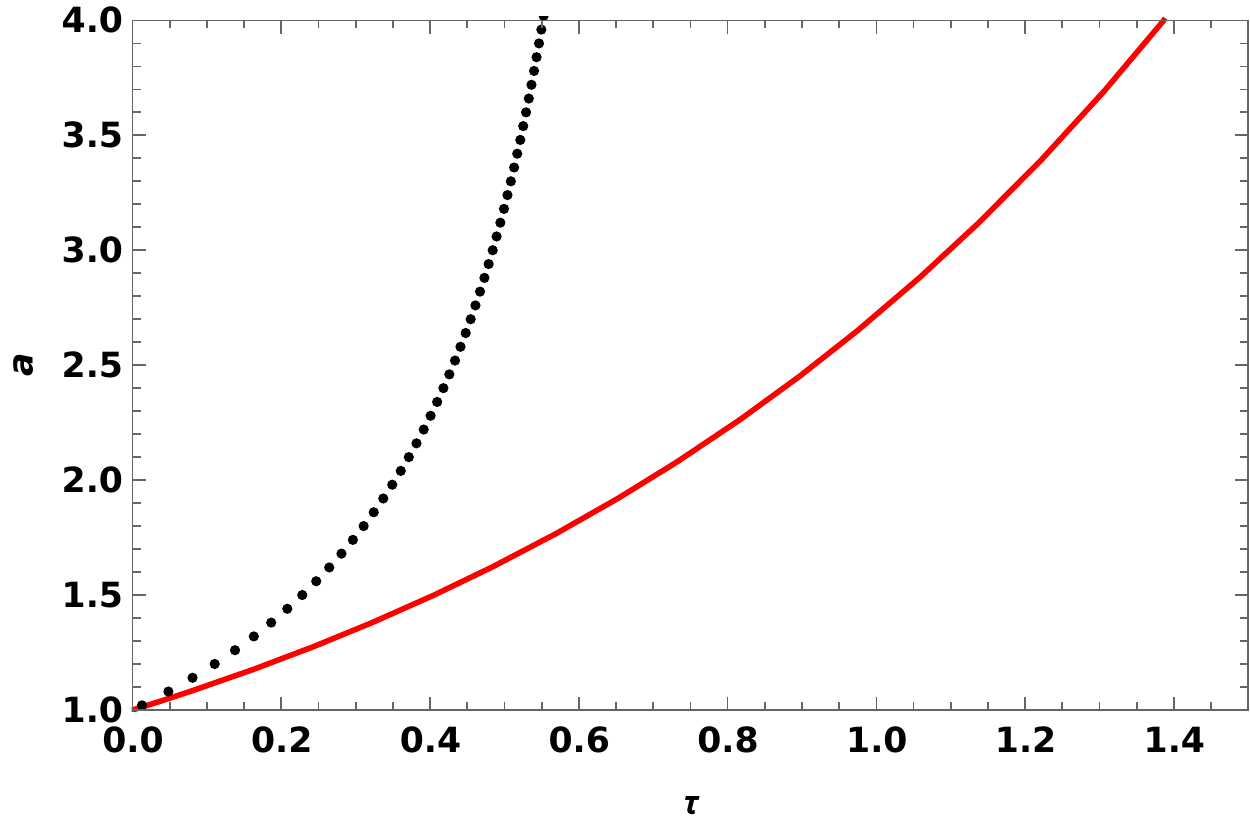}
\caption{Scale factor versus $\tau$ for cosmological constant. Solid red curve: classical scale factor. Dashed black curve: corrected scale factor. }
\label{fig4}
\end{figure}

\noindent In this subsection, we  will consider the Newtonian universe to be dominated by the cosmological constant. This case is particularly interesting as the role of the cosmological constant has not been investigated earlier in the Newtonian framework. Setting $\gamma=0$ in eq.\eqref{11}, we get
\begin{align}
\left(\frac{\mathrm{d} R}{\mathrm{d} t}\right)^{2}&=2c^{2}\phi\left(\frac{R_{0}}{R}\right)^{-2}\left[1+\lambda \phi \left( \frac{R_{0}}{R}\right)^{-2}- \zeta \left(\frac{l_{Pl}}{R_{0}}\right)^{2} \left(\frac{R_{0}}{R}\right)^{2}\right] \nonumber \\ & \equiv d_{1}\left(\frac{R}{R_{0}}\right)^{2}+d_{2}\left(\frac{R}{R_{0}}\right)^{4}-d_{3}~
\label{19}
\end{align}
where $d_{1}=2c^{2}\phi$, $d_{2}=2c^{2}\lambda \phi^{2}$ and $d_{3}=-2 \zeta c^{2}\phi \left(\frac{l_{Pl}}{R_{0}}\right)^{2}$. 
Integrating the above equation, we obtain the exact solution of the cosmological time in terms of the radius of our universe
\begin{align}
\frac{2\sqrt{d_{1}} }{R_{0}}t=\ln \left[\frac{1-\sqrt{1+\frac{d_{2}}{d_{1}}\left(\frac{R}{R_{0}}\right)^{2}}}{1+\sqrt{1+\frac{d_{2}}{d_{1}}\left(\frac{R}{R_{0}}\right)^{2}}} \right]+\frac{d_{3}d_{2}}{ d_{1}^{2}}\frac{_2F_1\left(-\frac{1}{2},2,\frac{1}{2};1+\frac{d_{2}}{d_{1}}\left(\frac{R}{R_{0}}\right)^{2}\right)}{\sqrt{1+\frac{d_{2}}{d_{1}}\left(\frac{R}{R_{0}}\right)^{2} }}+D
\label{N20}
\end{align}
where $D$ is the integration constant. We set the initial condition $R(t = t_{0}) = R_{0}$ as before. This immediately fixes the integration constant $D$ to be
\begin{align}
D=\ln \left[\frac{1-\sqrt{1+\frac{d_{2}}{d_{1}}}}{1+\sqrt{1+\frac{d_{2}}{d_{1}}}} \right]+\frac{d_{3}d_{2}}{ d_{1}^{2}}\frac{_2F_1\left(-\frac{1}{2},2,\frac{1}{2};1+\frac{d_{2}}{d_{1}}\right)}{\sqrt{1+\frac{d_{2}}{d_{1}}}}-\frac{2\sqrt{d_{1}} }{R_{0}}t_{0}~.
\label{consD}
\end{align}
The solution in terms of the scale factor then reads 
\begin{align}
\frac{2\sqrt{d_{1}}}{R_{0}} (t-t_{0}) &= \ln \left[\frac{1-\sqrt{1+\frac{d_{2}}{d_{1}}a^{2}}}{1-\sqrt{1+\frac{d_{2}}{d_{1}}}} \frac{1+\sqrt{1+\frac{d_{2}}{d_{1}}}}{1+\sqrt{1+\frac{d_{2}}{d_{1}}a^{2}}} \right] + \frac{d_{3}d_{2}}{d_{1}^{2}}\frac{_2F_1\left(-\frac{1}{2},2,\frac{1}{2};1+\frac{d_{2}}{d_{1}} a^{2}\right)}{\sqrt{1+\frac{d_{2}}{d_{1}}a^{2}}} \nonumber \\& \qquad \qquad -\frac{d_{3}d_{2}}{ d_{1}^{2}}  \frac{_2F_1\left(-\frac{1}{2},2,\frac{1}{2};1+\frac{d_{2}}{d_{1}}\right)}{\sqrt{1+\frac{d_{2}}{d_{1}}}}.
\label{dark_sol}
\end{align}
The classical solution for the cosmological constant from eq.\eqref{dark_sol}, setting $\lambda=0$ and $\zeta=0$ and thus $d_{2}=0$ and $d_{3}=0$ , is obtained to be
\begin{align}
a=\exp \left ( \tau \right )
\label{classdust}
\end{align}
where $\tau = \frac{\sqrt{d_{1}}}{R_{0}} (t-t_{0})$.\\
The plots of the scale factor vs. $\tau$ are illustrated in Fig.\eqref{fig4}. Here we have also considered $\rho_{0} = \rho_{c,0}$ as earlier so that once again the constant $\frac{d_{2}}{d_{1}}=\lambda \phi \approx 1.44$ for the same values of $H_{0}$, $\rho_{0}$ and $R_{0}$ mentioned earlier.
%
\section{Conclusions}
In this paper, we have investigated Newtonian cosmology taking into account the leading correction terms, namely, the leading classical term and quantum correction term to the Newtonian potential. We should point out that the phrase `Newtonian cosmology' does not refer to any special kind of cosmology, but to the fact that Newtonian gravity is used for calculating the expansion rate. As noted in~\cite{kour, Callan_2005}, it is not a coincidence that we get sensible results, but ``because the Newtonian theory of gravity is the weak-field limit of general relativity.'' Since quantum corrections to the Newtonian potential between two masses have been calculated, we decided to look at how the expansion of the Universe is affected. 

In this work, quantum Friedmann equations have been derived starting from the energy conservation of a test mass $m$ in a quantum corrected Newtonian potential. However, the leading $\mathcal{O}(\hbar)$ correction is negligibly small compared to the ``classical'' correction term of $\mathcal{O}(G^{2})$ which appears from a loop calculation. We have studied the cosmology for a single component universe, that is when the energy momentum tensor can be ascribed to radiation, dust, or the cosmological constant. We have calculated and plotted the solutions for the respective scale factors against the cosmological time. We have found that the evolution of the Newtonian universe in the presence of the leading classical term of order $\mathcal{O}(G^{2})$ term is similar for the radiation and dust cases at late times. However, the evolution of the Newtonian universe is quite different in the case of the cosmological constant at late times showing a steep increase in the scale factor.  
Another point to notice is that the Hubble parameter can vanish for radiation depending on the sign of the parameter $\lambda$ of the classical correction  term  no matter what the sign of the quantum corrected parameter takes.

\end{document}